# A new approach to the inverse problem for current mapping in thin-film superconductors

J. W. Zuber, F. S. Wells, S. A. Fedoseev, T. H. Johansen, A. B. Rosenfeld, and A. V. Pan

**Articles you may be interested in**

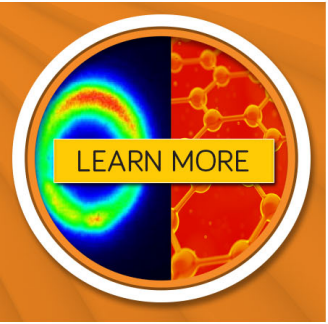

# A new approach to the inverse problem for current mapping in thin-film superconductors

J. W. Zuber,[1] F. S. Wells,[1] S. A. Fedoseev,[1,2] T. H. Johansen,[1,3] A. B. Rosenfeld,[2] and A. V. Pan[1,4,a)]

[1]*Institute for Superconducting and Electronic Materials, University of Wollongong, Northfields Avenue, Wollongong, NSW 2522, Australia*
[2]*Centre for Medical Radiation Physics, University of Wollongong, Northfields Avenue, Wollongong, NSW 2522, Australia*
[3]*Department of Physics, University of Oslo, Blindern, 0316 Oslo, Norway*
[4]*National Research Nuclear University MEPhI (Moscow Engineering Physics Institute), 31 Kashirskoye Shosse, 115409 Moscow, Russian Federation*

A novel mathematical approach has been developed to complete the inversion of the Biot-Savart law in one- and two-dimensional cases from measurements of the perpendicular component of the magnetic field using the well-developed Magneto-Optical Imaging technique. Our approach, especially in the 2D case, is provided in great detail to allow a straightforward implementation as opposed to those found in the literature. Our new approach also refines our previous results for the 1D case [Johansen *et al.*, Phys. Rev. B **54**, 16264 (1996)], and streamlines the method developed by Jooss *et al.* [Physica C **299**, 215 (1998)] deemed as the most accurate if compared to that of Roth *et al.* [J. Appl. Phys. **65**, 361 (1989)]. We also verify and streamline the iterative technique, which was developed following Laviano *et al.* [Supercond. Sci. Technol. **16**, 71 (2002)] to account for in-plane magnetic fields caused by the bending of the applied magnetic field due to the demagnetising effect. After testing on magneto-optical images of a high quality YBa$_2$Cu$_3$O$_7$ superconducting thin film, we show that the procedure employed is effective. *Published by AIP Publishing.*
https://doi.org/10.1063/1.5012588

## I. INTRODUCTION

Magnetic imaging techniques are increasingly prevalent in many fields of science and medicine, as they can be used to determine the properties of materials and provide powerful medical diagnostic tools through non-contact, non-destructive measurements. Such techniques in physics and materials science research include but are not limited to Magnetic force microscopy;[1] SQUID (Superconducting Quantum Interference Device) microscopy,[2,3] as well as the more recent invention of SQUID-on-tip microscopy, whereby a nanoSQUID is fabricated on a quartz tip;[4,5] Magneto-optical imaging (MOI) for ferromagnetic and superconducting materials,[6,7] and its dynamic version for rapid imaging of transient current effects;[8,9] μHall-probe microscopy;[6,11–14] bitter decoration;[15] scanning magnetoresistive microscopy;[16] scanning electron microscopy with polarization analysis (SEMPA);[17] and electron holography.[18]

In medical and biological applications, such techniques include optical magnetic imaging of living cells,[19] magnetoencephalography (MEG) of brain activity,[20–22] magnetocardiography (MCG) for magnetic heart signals,[20,23] and Magnetic Resonance Imaging (MRI) for producing images of soft tissues, organs, body, and brain.[24] MRI has become a valuable technique for medical diagnostics and biomedical research with a large number of important developments, such as functional Magnetic Resonance Imaging (fMRI) of human brains[25,26] and Magnetic Resonance Elastography measuring the mechanical properties of soft tissues of the brain.[27]

The MEG/MCG image reconstruction is largely based on the solution to the inverse problem.[20] The similar inverse problem of the Biot-Savart law also allows the calculation of plane current components from magnetic field measurements of a sample. This is particularly useful for superconductors and thin films,[7–10,28] ferromagnetic and superconducting hybrids,[29,30] and even magnetic flux quanta (Abrikosov vortices).[3]

In practice, methods of magnetic field detection via magnetometers often measure only the out-of-plane z-component of the magnetic field extending from a sample. This adds complexity to the inversion problem since two current components, $J_x$ and $J_y$, must be determined from only one field component, $B_z$. Several solutions to this problem have been proposed over the years. However, the conditions of applicability have varied. For example, Roth *et al.*[33] solved the problem considering that the magnetic field sensor was placed at a finite distance from the sample which is much larger than the sample thickness, which considerably simplified the approach but reduced its accuracy. Johansen *et al.* initially solved the non-contact problem only in a long strip 1D case.[31] A more general solution to the problem was devised by Jooss *et al.*, which takes into account the finite distance to the detector, the sample thickness, and solves the problem in 2D.[32]

A new and simpler method for solving the inverse problem is described in this paper, and the final result is consistent with Jooss' solution. This new approach requires neither Green's function integral identities[32] nor Topelitz matrices

a)Author to whom correspondence should be addressed: pan@uow.edu.au

for numerical solutions to the Biot-Savart law.[34] Instead, computation of the integral kernel is carried out directly by using properties of the Bessel functions of the first kind and the Laplace transform.

Though a magnetic field is applied solely in the z-direction, in-plane magnetic fields exist at the finite measurement distance in MOI due to the bending of magnetic fields around the superconducting sample by the demagnetising effect. To account for this, a technique was proposed by Johansen et al.,[35] and later described by Laviano et al.[36] for the 2D case. This 2D technique involves computing the x and y components of the magnetic field from the calculated current in the film, and then re-calculating the currents from the new field distribution in an iterative procedure. This procedure was streamlined and verified by testing it on magneto-optical images of high quality $YBa_2Cu_3O_{7-\delta}$ (YBCO) thin films. It appears to be effective, but may produce noticeable artefacts on the corrected images of field and current if over-applied due to the internal LabVIEW Fast Fourier Transform (FFT) routine used in our iterations.

The accuracy of any correction technique is vital to the practice of MOI, since it can avoid calculation of misleading current distributions. Such techniques could also lead to higher resolution image reconstructions for determination of local electrodynamic quantities, allow more accurate determination of properties such as local current-carrying capacity,[7,31,33,36,37] and refine MEG/MCG image reconstruction for the human brain in response to demand for single-cell resolution.[38–40]

## II. THEORY

In detecting the magnetic field around superconducting samples, the magneto-optical imaging technique measures the local z-component of the magnetic field at each point in a Faraday-active indicator film,[9] which is placed at a finite distance h above the sample. In our experimental setup, h is taken to be the distance between the top of the sample and the bottom of the indicator film (Fig. 1) and is estimated from the sample and film roughness, and the thickness of the reflective layer on the indicator film as also considered in.[8,28,31,32] Furthermore, there also exist in-plane fields ($B_{xy}$), which contribute to the currents created within the sample. The indicator film is also sensitive to these field components. Such in-plane fields are caused by the bending of the imposed magnetic field around the sample due to the demagnetising effect, and due to the stray fields from each vortex bending outward as they reach the height of the indicator film.

Figure 1 schematically shows a thin superconducting sample with a magneto-optical indicator film on top of it. An applied magnetic field bends around the superconductor due to screening of the magnetic field. This demagnetising effect is shown by the curved arrows on both sides of the sample in Fig. 1(a). A vortex has exponentially diverging stray fields above the sample[12,41] [shown schematically by arrows below and above the vortex in Fig. 1(a)], which overlap for a large number of vortices at the distance where the indicator film is located. As a result, the in-plane field $B_{xy}$ and out-of-plane field $B_z$ components exist at the measurement height since the direction of the magnetic field lines are at some angle to the z-axis.

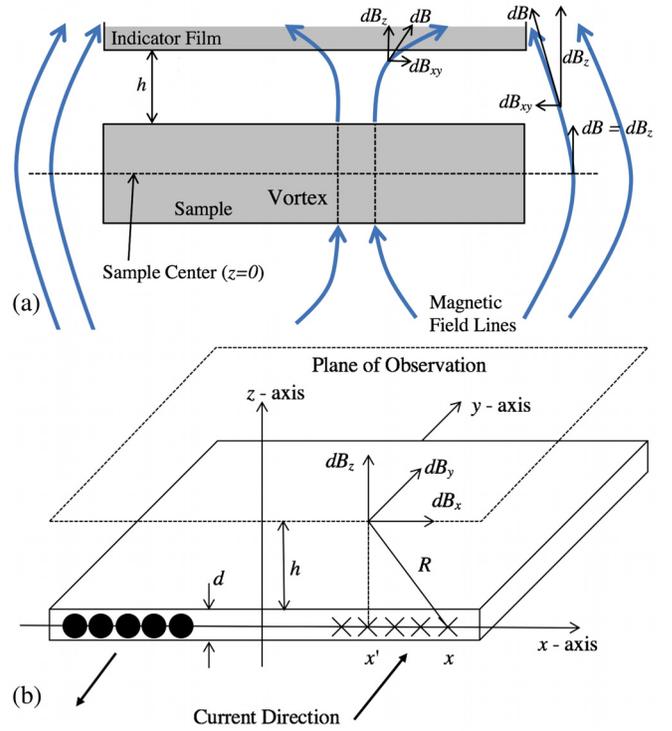

FIG. 1. (a) Bending of the imposed magnetic field (bold curved arrows) around a superconducting sample due to the demagnetising effect and spreading of the stray field above vortices. (b) Visualisation of the inverse problem in one dimension.

When a magnetic field is present at a point on the indicator film, the local magnetic moment is perturbed by an angle $\phi$ as shown in Fig. 2. Consideration of the geometry in Fig. 2 allows us to write the interaction energy, $E_{int}$, of the indicator film in the presence of an applied magnetic field with a non-zero z-component as in Ref. 36, Eq. (3.1)

$$E_{int} = E_A(1 - \cos\phi) + BM_s[1 - \cos(\alpha - \phi)], \quad (1)$$

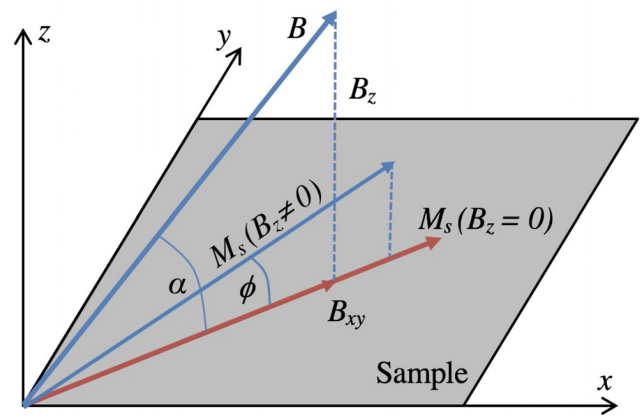

FIG. 2. When there is no applied field in the z-direction, the spontaneous magnetisation vector lies in the plane of the film. A non-zero z-component of the applied field perturbs the magnetisation vector by an angle $\phi$, giving it a non-zero z-component.

where $E_A$ is the anisotropy energy, $M_s$ is the value of the spontaneous magnetisation, and $\alpha$ is the angle formed between the magnetic induction **B** and the *xy*-plane.

To find an expression for the equilibrium magnetisation, an angle $\phi$ is determined such that the interaction energy is a minimum. This angle represents the balance between the magnetocrystalline anisotropy and the tendency to align with the external magnetic field.[31] Hence, computing the derivative of Eq. (1) with respect to $\phi$ and showing it has second derivative always positive, gives

$$\phi = \tan^{-1}\left(\frac{B_z}{B_A + B_{xy}}\right), \quad (2)$$

defining the following: $\frac{E_A}{M_s} = B_A$ as the magnetic anisotropy field and $B_{xy} = \sqrt{(B \cdot \hat{x})^2 + (B \cdot \hat{y})^2} = B\cos\alpha$.

The polarisation of light incident on the indicator film is rotated by the Faraday effect. In the MOI setup, polarised light is applied to the indicator film parallel to the film's optical axis. Therefore, the angle of rotation $\alpha_F$ is proportional to the magnetisation along the light propagation direction, allowing us to use the Faraday rotation law

$$\alpha_F = CM_s \sin\phi, \quad (3)$$

where $C$ is an experimental constant dependent on the indicator film thickness. After rotation the light passes through an analyser at a fixed angle $\theta$ with respect to the polariser. Invoking Malus' law then gives the intensity at the analyser as

$$I = I_0 + I_{\max} \cos^2(\alpha_F + \theta). \quad (4)$$

The light intensity is then measured employing a camera, and the magnetic field can be determined directly from the measured light intensity. This light intensity can be used to calculate currents in a superconducting film placed underneath (Fig. 1). We consider this problem quantitatively in the Subsections III A and III B.

However, it is worth mentioning that the magnetic field varies along the *z*-axis due to the magnetic field bending as shown by the field lines [bold curved arrows in Fig. 1(a)]. In particular, within the indicator film, the light intensity becomes an integral quantity over the indicator film thickness, commonly ranging from a few hundred nanometers to a few microns.[42] This effect has never been taken into account and is also not considered in our work due to a somewhat arbitrary choice of the parameter $h$ as discussed above. This choice ultimately makes any quantitative assessment of this integral quantity carry some degree of uncertainty, resulting in some error in the absolute values of calculated currents. This is a limitation of this type of analysis. However, the impact on the results can be minimised by ensuring the smallest possible $h$ and employing as thin indicator films as practical.

## III. THE INVERSE PROBLEM

The inverse problem is the problem of calculating currents from magnetic field measurements of a thin film superconductor placed in a homogeneous perpendicular magnetic field. Since magnetic field values are used to calculate current in a non-local manner, the Biot-Savart law is always taken as the starting point of the problem.

### A. The inverse problem in 1D

The solution to the 1D inverse problem depicted in Fig. 1 has been done Johansen *et al.*[31] We re-derive it here with a somewhat different approach and more details. It is to allow a simplified practical implementation for research, visualisation, or quality control purposes.

The sheet is taken to be thin so that $J_z = 0$, and since we are only considering the 1D case, $J_x = 0$ here too. For the currents, the sheet thickness, $d$, is taken to be small compared to the height of the detector above the sheet, $h$: $d \ll h$. Hence, in the application of the Biot-Savart law, the sheet current is given by

$$J_y(x) = \int_0^d j(x,z)dz, \quad (5)$$

where $j(x, z)$ is the local current density caused by current moving in the *y*-direction producing a magnetic field in the *xz*-plane.

An application of the Biot-Savart law in 1D gives

$$B_z(x') = \frac{\mu_0}{2\pi}\int_{-\infty}^{\infty}\frac{x-x'}{h^2+(x-x')^2}J(x)dx. \quad (6)$$

Prior to taking the Fourier transform of the inhomogeneous Fredholm integral equation of the first kind above observe that the kernel

$$K(x,x') = \frac{x-x'}{h^2+(x-x')^2} = K(x-x'), \quad (7)$$

is translational invariant, so a convolution in the *x*-plane can be applied. Hence, the Fourier transform of Eq. (7) where $x' \to k_x$ yields

$$\tilde{B}_z(k_x) = \frac{\mu_0}{2\pi}\tilde{K}(k_x)\tilde{J}(k_x). \quad (8)$$

Now, obtaining an expression for $J(x)$—the current value at the small region of current not the *x*-ordinate of the magnetometers position $(x')$—and using the inversion theorem gives

$$\mu_0 J(x) = \int_{-\infty}^{\infty}\frac{\tilde{B}_z(k_x)}{\tilde{K}(k_x)}e^{ik_x x}dk_x. \quad (9)$$

In order to compute the Fourier transform of the integral kernel, it can be shown

$$\tilde{K}(k_x) = \int_{-\infty}^{\infty}\frac{x'}{h^2+x'^2}e^{-ik_x x'}dx' = -i\pi\mathrm{sgn}(k_x)e^{-h|k_x|}, \quad (10)$$

where $\mathrm{sgn}(x)$ is the sign function. A proof is provided in Appendix A.

Before discretisation of the integral can take place, the functions in the integrand must be transformed into those which can be represented by an infinite series. So, using Eq. (10)

$$\mu_0 J(x) = \int_{-\infty}^{\infty} \frac{\tilde{B}_z(k_x)}{-i\pi \text{sgn}(k_x) e^{-h|k_x|}} e^{ik_x x} dk_x. \quad (11)$$

However, noting that the transfer function is given by $\frac{1}{\tilde{K}(k_x)}$, it is evident that components with high frequency will be highly amplified, since by Eq. (10)

$$\frac{1}{\tilde{K}(k_x)} \propto e^{h|k_x|}. \quad (12)$$

Such components are to be removed. Hence, a low pass filter should be included in the analysis so that components with $|k_x| \geq K_c$ are cut out, where $K_c$ is a cut-off frequency. Including this cut-off frequency gives

$$\mu_0 J(x) = \int_{-\infty}^{\infty} \int_{-K_c}^{K_c} \frac{e^{ik_x x} B_z(x') e^{-ik_x x'}}{-i\pi \text{sgn}(k_x) e^{-h|k_x|}} dk_x dx'. \quad (13)$$

By defining

$$A(\xi) = \int_{-K_c}^{K_c} \frac{e^{ik_x \xi}}{-i\pi \text{sgn}(k_x) e^{-h|k_x|}} dk_x, \quad (14)$$

Eq. (13) may be written as

$$\mu_0 J(x) = \int_{-\infty}^{\infty} A(x - x') B_z(x') dx'. \quad (15)$$

The following solution to Eq. (14) is given in Ref. 31:

$$A(\xi) = \frac{\xi [1 - e^{k_c h} \cos(K_c \xi)] + h e^{K_c h} \sin(K_c \xi)}{h^2 + \xi^2}; \quad (16)$$

however, it lacks a small numerical factor of $-\frac{2}{\pi}$, which we re-introduce after careful mathematical analysis, as follows:

$$A(\xi) = \int_{-K_c}^{K_c} \frac{e^{ik_x \xi}}{-i\pi \text{sgn}(k_x) e^{-h|k_x|}} dk_x$$

$$= \frac{i}{\pi} \left[ \int_{-K_c}^{0} \frac{e^{k_x(i\xi - h)}}{-1} dk_x + \int_{0}^{K_c} \frac{e^{k_x(h + i\xi)}}{1} \right]$$

$$= \frac{-i}{\pi} \left[ \frac{i\xi e^{-K_c(i\xi - h)} - i\xi + h e^{-K_c(i\xi - h)}}{h^2 + \xi^2} \right.$$

$$\left. + \frac{i\xi e^{K_c(i\xi + h)} - i\xi - h e^{K_c(i\xi + h)}}{h^2 + \xi^2} \right]$$

$$= \frac{2}{\pi} \frac{\xi [e^{K_c h} \cos(\xi K_c) - 1] - h e^{K_c h} \sin(\xi K_c)}{h^2 + \xi^2},$$

$$\therefore A(\xi) \neq \frac{\xi [1 - e^{k_c h} \cos(K_c \xi)] + h e^{K_c h} \sin(K_c \xi)}{h^2 + \xi^2}. \quad (17)$$

With such an integral kernel, $A(\xi)$, the integral in Eq. (15) is discretised since magnetic flux is measured in discrete pixel-sized units. For this process, $\Delta = \pi / K_c$ was defined as the unit length, then the x-coordinates were discretised as: $x = n\Delta$ and $x' = n'\Delta$, and finally $h = t\Delta$, where $t$ is simply the film thickness under the change of variables.

The infinite sum representing the revised version of Eq. (15) was calculated to be

$$\mu_0 J(n) = -\frac{2}{\pi} \sum_{n'} \frac{n - n'}{t^2 + (n - n')^2} \frac{1 - (-1)^{n - n'} e^{\pi t}}{\pi} 2B_z(n'). \quad (18)$$

Hence, the 1D inverse problem has been solved. A discrete sum is obtained here to represent the current density given the magnetic field values. However, this solution was not consistent with previous literature due to the aforementioned small factor of $-\frac{2}{\pi}$.

Therefore, the current appears to flow in the opposite direction and has a slightly smaller amplitude by a factor of $2/\pi$, compared to the result in Ref. 31. The true "Current Direction" is shown in Fig. 1(b) along the x-axis.

### B. The inverse problem in 2D

The inverse problem in two dimensions was analysed using a novel approach, which somewhat resembles that given by Jooss *et al.*[32] It is however unique, since instead of using Green's functions to evaluate the Fourier transform of the integral kernel, it is represented as a Hankel transform for which the Bessel function and Laplace transform identities can then be used to compute the transform.

To derive a formula for $J_x(x, y)$ and $J_y(x, y)$ as a function of $B_z$, we start with the continuity condition for the sheet current density

$$\nabla \cdot \mathbf{J}(x, y) = 0. \quad (19)$$

It assumes that there is no current flowing in the z-direction of the sheet since it is thin.

We then define a scalar function $g(x, y)$ by the relation

$$\mathbf{J}(x, y) = \nabla \times \hat{z} g(x, y), \quad (20)$$

which is consistent with the continuity equation (19). To alleviate the freedom of gauge choice, the Coulomb gauge condition can be applied so that $g(x, y)$ is totally defined as the local magnetisation and is the potential function for the current density.

Using the new expression for $\mathbf{J}(x, y)$ in the integrand given by the Biot-Savart law, we obtain

$$B_z(\mathbf{r}) = \frac{\mu_0}{4\pi} \int_V \hat{z} \frac{[\nabla \times \hat{z} g(x, y)] \times (\mathbf{r} - \mathbf{r}')}{|\mathbf{r} - \mathbf{r}'|^3} d^3 \mathbf{r}', \quad (21)$$

where $\mathbf{r}$ is the position we measure the field being produced from position $\mathbf{r}'$ and the volume element $d^3 \mathbf{r}' = dx' dy' dz'$. Note once again that we are only able to measure the z-component of the magnetic field, hence the unit vector in the integrand.

Now, it is left to compute the integrand in a form, where the convolution theorem for Fourier transforms can be applied. It can be shown using vector calculus identities that

$$B_z(\mathbf{r}) = -\frac{\mu_0}{4\pi} \int_V \frac{2(z - z')^2 - (x - x')^2 - (y - y')^2}{\left[(x - x')^2 + (y - y')^2 + (z - z')^2\right]^{\frac{5}{2}}} g(x', y') d^3 \mathbf{r}'. \quad (22)$$

Taking into account the finite thickness of the sample ($d$), we integrate $z'$ over the domain $\frac{-d}{2}$ to $\frac{d}{2}$. Now in taking the Fourier transform of the $x'$ and $y'$ integrals (since we may swap the order of integration so as to only transform the inner integrals, then integrate the result over the $z'$ domain); mapping the spatial variables $x$ and $y$ to the frequency variables $k_x$ and $k_y$, respectively; applying the convolution theorem in the $xy$-plane; and noting that we are always measuring the field at a constant height $h$ above the surface of the sample, hence $z = h$, we obtain

$$\tilde{B}_z(k_x, k_y, h, d) = -\frac{\mu_0}{4\pi} \int_{-\frac{d}{2}}^{\frac{d}{2}} \\ \times \mathcal{F}\left\{\frac{2(h-z')^2 - x^2 - y^2}{\left[x^2 + y^2 + (h-z')^2\right]^{\frac{5}{2}}}\right\} \tilde{g}(k_x, k_y) dz'. \quad (23)$$

To compute the Fourier transform shown above, we claim that

$$\mathcal{F}\left\{\frac{2(h-z')^2 - x^2 - y^2}{\left[x^2 + y^2 + (h-z')^2\right]^{\frac{5}{2}}}\right\} = 2\pi k e^{-k(h-z')}, \quad (24)$$

where $k = \sqrt{k_x^2 + k_y^2}$. This claim may be re-written using the inversion theorem as

$$\frac{2(h-z')^2 - x^2 - y^2}{\left[x^2 + y^2 + (h-z')^2\right]^{\frac{5}{2}}} = 2\pi \mathcal{F}^{-1}\left[k e^{-k(h-z')}\right]. \quad (25)$$

The proof of the claim is given in Appendix B.
Hence, the magnetic field detected becomes

$$\tilde{B}_z(k_x, k_y, d, h) = -\frac{\mu_0}{4\pi} \int_{-\frac{d}{2}}^{\frac{d}{2}} e^{-k(h-z')} 2\pi k \tilde{g}(k_x, k_y) dz'. \quad (26)$$

Now, we can assume that $\tilde{g}(k_x, k_y)$ is not dependent on $z'$ because we assume that the current only flows in the $xy$-plane

$$\tilde{B}_z(k_x, k_y, d, h) = -\frac{\mu_0}{2} \tilde{g}(k_x, k_y) k e^{-kh} \int_{-\frac{d}{2}}^{\frac{d}{2}} e^{kz'} dz'$$
$$= -\mu_0 \tilde{g}(k_x, k_y) e^{-kh} \sinh\left(\frac{kd}{2}\right). \quad (27)$$

The only subsequent analysis to be performed is to transform Eq. (27) into a form to which an inverse Fast Fourier Transform (FFT) algorithm can be applied resulting in a current map of the sample. We make use of the definition of the function $g(x, y)$

$$\mathbf{J}(x, y) = \nabla \times \hat{z} g(x, y) = [\partial_y g(x, y), -\partial_x g(x, y), 0], \quad (28)$$

$$\tilde{J}_x(k_x, k_y) = -i k_y \tilde{g}(k_x, k_y), \quad (29)$$

$$\tilde{J}_y(k_x, k_y) = i k_x \tilde{g}(k_x, k_y), \quad (30)$$

and the continuity equation

$$k_x \tilde{J}_x(k_x, k_y) + k_y \tilde{J}_y(k_x, k_y) = 0. \quad (31)$$

Finally, we can obtain expressions for both components of the current as functions of physical variables and the magnetic field detected in the $z$-direction by using Eqs. (27), (29), and (30)

$$\tilde{J}_x(k_x, k_y) = \frac{ik_y}{\mu_0} e^{kh} \operatorname{cosech}\left(\frac{kd}{2}\right) \tilde{B}_z(k_x, k_y, h, d), \quad (32)$$

$$\tilde{J}_y(k_x, k_y) = \frac{-ik_x}{\mu_0} e^{kh} \operatorname{cosech}\left(\frac{kd}{2}\right) \tilde{B}_z(k_x, k_y, h, d). \quad (33)$$

Both of the above equations can have the FFT algorithm applied to them so that the values of $B_z$ obtained can be used to calculate $J_x$ and $J_y$ and hence obtain a current map of the superconductor. Thus, solving the inverse problem and ending with final equations consistent with those derived by Jooss *et al.*[32]

Another method for solving the inverse problem in 2D was developed by Roth *et al.*[33] In contrast to the method considered above, this method does not take into account the finite height of the magnetometer above the sample; hence, it assumes detection of the magnetic field directly at the sample surface. Mathematically, it differs in that the integral kernel after a Fourier transform is computed directly from the Biot-Savart law for all components of the magnetic field. Since this method does not take into account typical experimental conditions, it is considered to be inaccurate. Hence, our method described above is employed for subsequent analysis in this work.

## IV. IN-PLANE CORRECTION

The intensity at the analyser in the equilibrium position in a magneto-optical imaging apparatus is found by substituting Eqs. (2) and (3) into (4), the result gives

$$I = I_0 + I_{\max} \cos^2\left[\frac{CM_s B_z}{\sqrt{(B_A + B_{xy})^2 + B_z^2}} + \theta\right]. \quad (34)$$

It has been common practice to take $B_{xy} = 0$ for simplicity, however, that leads to an interpretation of MOI data that may not be physically accurate.[32,33] Such an assumption causes an un-physically higher electrical current (increasing with sample thickness) to be observed.[36]

Therefore, the in-plane correction procedure was proposed[36] in order to form a relationship between the apparent magnetic field at the detector assuming $B_{xy} = 0$ (henceforth written as $B_{z|0}$) and the field at the detector including the in-plane effects.

Below, we scrutinize the in-plane correction proposed by Laviano *et al.*[36]

Using Eq. (34), we write an expression for $B_z$

$$\frac{B_z}{B_A + B_{xy}} = \tan\left\{\sin^{-1}\left[\frac{\cos^{-1}\left(\sqrt{\frac{I - I_0}{I_{\max}}}\right) - \theta}{CM_s}\right]\right\}. \quad (35)$$

With the assumption that $B_{xy} = 0$, this equation becomes

$$\frac{B_{z|0}}{B_A} = \tan\left\{\sin^{-1}\left[\frac{\cos^{-1}\left(\sqrt{\frac{I-I_0}{I_{max}}}\right)-\theta}{CM_s}\right]\right\}. \quad (36)$$

Combining Eqs. (35) and (36) gives

$$B_z = \left[1 + \frac{\sqrt{B_x^2 + B_y^2}}{B_A}\right] B_{z|0}. \quad (37)$$

This neatly presented equation is used in an attempt to produce better approximations of $B_z$ from known $B_{z|0}, B_x,$ and $B_y$ values while taking $B_A$ as a constant. It was proposed[36] that an iterative procedure be carried out, with corrected values of magnetic field at every point on the sample determined by applying the following algorithm:

(i) Use $B_z^{(n)}$ to find $J_x^{(n)}(x,y)$ and $J_y^{(n)}(x,y)$ using an FFT algorithm on the Eqs. (32) and (33), and starting with $B_z^{(0)} = B_{z|0}$;
(ii) Calculate $B_x^{(n)}(x,y)$ and $B_y^{(n)}(x,y)$ from $J_x^{(n)}(x,y)$ and $J_y^{(n)}(x,y)$;
(iii) Calculate $B_z^{(n+1)}$ using

$$B_z^{(n+1)} = \left[1 + \frac{\sqrt{B_x^{(n)2} + B_y^{(n)2}}}{B_A}\right] B_{z|0}, \quad (38)$$

(iv) Assign $B_z^{(n+1)} = B_z^{(n)}$, and go to (i), unless the difference between successive iterations is suitably small.

To complete stage (ii) of the algorithm proposed above, equations for $B_x$ and $B_y$ in terms of $B_{z|0}$, are determined using the inversion techniques previously presented.

By considering again the 2D inversion problem, but now solving for $B_x$ and taking into account the finite height of the detector

$$B_x(x,y,z,h) = \frac{\mu_0}{4\pi} \times \int_{-\frac{d}{2}}^{\frac{d}{2}}\int_{-\infty}^{\infty}\int_{-\infty}^{\infty} \\ \times \frac{J_y(x',y')(h-z')dx'dy'dz'}{\left[(x-x')^2 + (y-y')^2 + (h-z')^2\right]^{\frac{3}{2}}}, \quad (39)$$

which can be Fourier transformed with an application of the convolution theorem to gain an integral kernel with a derivative identical to the negative of that in (24). Hence, employing the claim in (24) then integrating gives

$$\tilde{B}_x(k_x,k_y,h) = \frac{\mu_0}{2\pi}\int_{-\frac{d}{2}}^{\frac{d}{2}} (h-z')\frac{e^{-\sqrt{k_x^2+k_y^2}(h-z')}}{h-z'}\tilde{J}_y(k_x,k_y)dz', \quad (40)$$

which can be integrated to find

$$\tilde{B}_x(k_x,k_y,d,h) = \frac{\mu_0 e^{-kh}}{k}\tilde{J}_y(k_x,k_y)\sinh\left(\frac{kd}{2}\right). \quad (41)$$

Similarly, for $\tilde{B}_y$

$$\tilde{B}_y(k_x,k_y,d,h) = -\frac{\mu_0 e^{-kh}}{k}\tilde{J}_x(k_x,k_y)\sinh\left(\frac{kd}{2}\right). \quad (42)$$

These are the expressions used in Ref. 36 for the calculation of $B_x$ and $B_y$, but simpler expressions may be found by substitution of Eq. (32) into (42) and (33) into (41). This leads to

$$\tilde{B}_x = \frac{ik_x}{k}\tilde{B}_z, \quad (43)$$

$$\tilde{B}_y = \frac{ik_y}{k}\tilde{B}_z. \quad (44)$$

These equations may be used in step (ii) of the algorithm to provide a streamlined correction procedure that remains analytically identical.

## V. EXPERIMENTAL VERIFICATION

This in-plane correction technique was applied to a number of magneto-optical images. The results are presented for a 3 mm² YBCO thin film. All images tested were acquired using the MOI technique and the apparatus described in Ref. 8. The sample is cooled to a temperature of 4 K in a Janis continuous-flow helium cryostat. The YBCO films used have been grown using the pulsed laser deposition technique.[43,44] Their typical $J_c \simeq 3 \times 10^{10}$ A/m² at $T = 77$ K and $T_c \simeq 91.0 \pm 0.5$ K.

After appropriate calibration, the grey-scale value of each pixel in the acquired MO images was converted to a magnetic field value. The resultant images, along with Eqs. (32) and (33), were used to calculate current maps. Finally, the correction procedure was applied, using the algorithm discussed in Sec. IV, taking the anisotropy field to be $B_A = 80$ mT.[31] The number of iterations of this technique was varied, and the effect on the resultant images was observed. Note that each Fourier transform was required to be discreet in the experimental case, as the measurements were discrete for each pixel.

The results of the correction procedure are presented in Figs. 3(b)–3(e). The desired reduction of currents at the edges of the sample is clearly seen in the disappearance of the bright border around the sample in the current images. This border is present in the right image of Fig. 3(a), but absent in all other current images [Figs. 3(b)–3(e)]. This is evidence that the effects of in-plane fields have been successfully removed.

However, an undesired effect has also emerged when the number of iterations is >5: a series of non-physical bright and dark fringes are seen close to the edges of the sample in both the field and current images [Figs. 3(d)–3(e)], with approximately one additional fringe arising with each iteration. These non-physical fringes are seen most clearly when no high-frequency filtering is applied during the current calculation procedure. The use of a Hanning filter with an empirically chosen cut-off frequency $K_c$, is commonly used to remove high-frequency noise from the calculated current maps[31,32] and smooths out these fringes. This is shown in Fig. 4, but it is clear [especially from parts (c) and (d) of this figure] that the influence of this artefact cannot be completely removed through filtering. The artefact appears and builds up only upon repetitive application of the iteration procedure starting from the 6th iteration. Each iteration step involves internal LabVIEW FFT and then inverse FFT routines, hence the artefact can be attributed to the discrete nature of these routines and associated binning factor (data quantisation) in numerical image processing. The binning factor of the centered Fourier spectrum gets larger and larger with each iteration step and would be expected to produce an ever increasing number of fringes with an ever larger spatial period, making them visible (exactly what occurs upon increasing the number of iterations). Smoothing the fringes upon filtering is another expected behaviour, again pointing at the problem with these FFT routines.

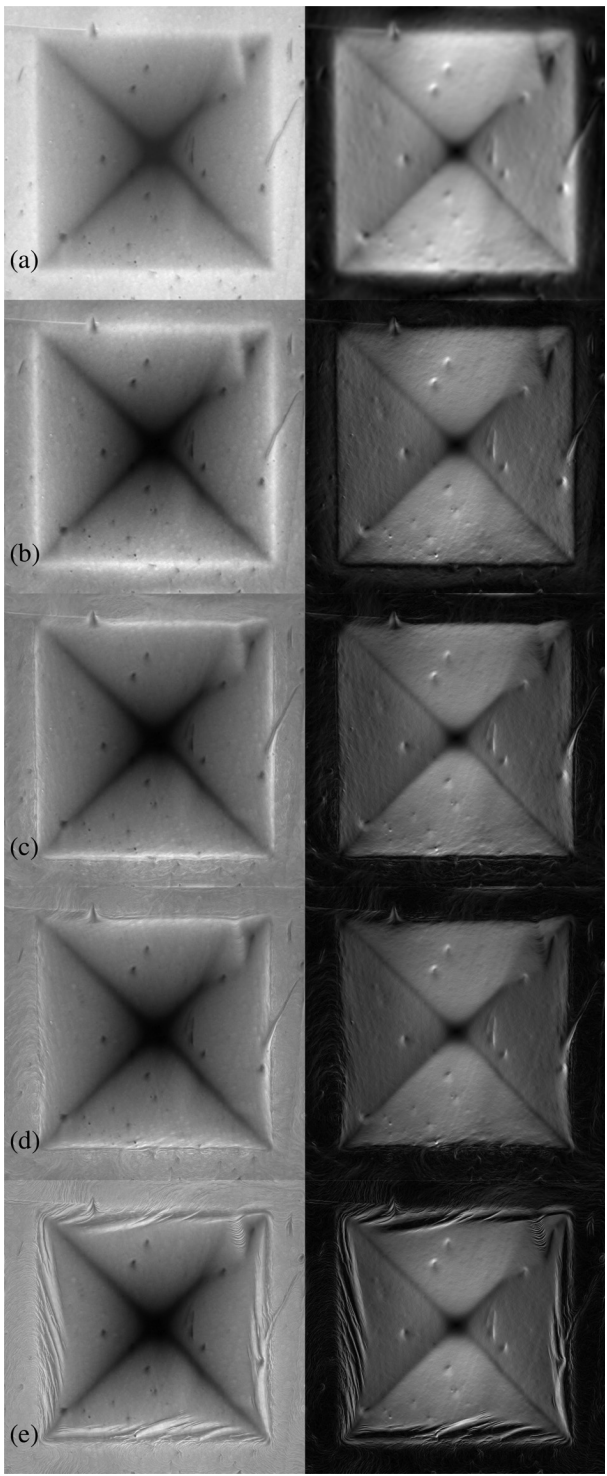

FIG. 3. Calculated values of the magnetic field (left) and current density (right) in the sample (a) without any correction (the left image is the original magneto-optical image), and after (b) 1, (c) 5, (d) 10, and (e) 100 iterations [Eq. (38)]. No high-frequency filtering is applied.

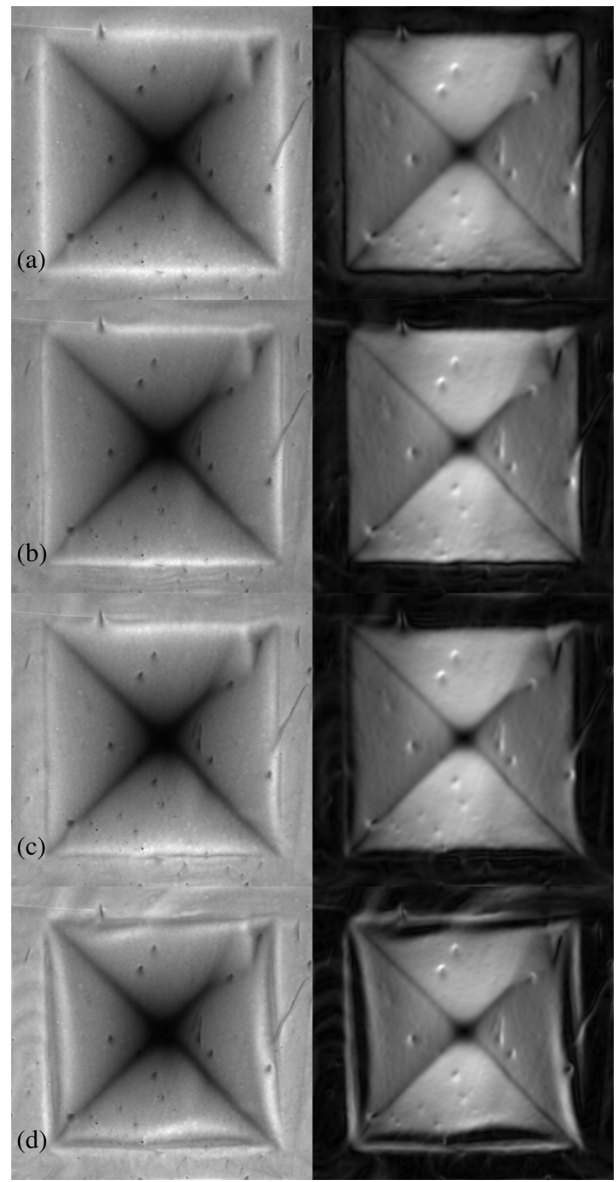

FIG. 4. Calculated values of the magnetic field (left) and current density (right) in the sample after (a) 1, (b) 5, (c) 10, and (d) 100 iterations [Eq. (38)], with high-frequency filtering using a Hanning window.

## VI. CONCLUSION

The inverse problem for finding the current distribution in superconducting thin films from magnetic field measurements in a plane above the film has been uniquely solved for both the one- and two-dimensional cases. A new method for computation of the integral kernel in the 2D case has been devised which involves reducing the problem to a Laplace transform of a Bessel function of the first kind. The derivations are provided in great detail to enable a straightforward implementation, as opposed to those available in the literature with limited information.

In the 2D case, the solution was found to be consistent with that previously determined in Refs. 32 and 33 although the use of Green's function identities were not required whilst still taking into account the finite height of the magnetisation above the surface.

In the 1D case, it was found that the calculated current appears to flow in the opposite direction and has a slightly smaller amplitude by a factor of $2/\pi$, compared to the result in Ref. 31. Otherwise, the solution is fully consistent with Ref. 31.

For the in-plane field correction procedure proposed in Ref. 36, the equations used to calculate the $x$ and $y$ components of the magnetic field at the detector [Eqs. (43) and (44)] were simplified slightly for our new approach, which is provided in every detail to enable straightforward implementation. The final equation and correction algorithm are fully consistent with the result obtained in Ref. 36.

Experimental verification of our approach is efficient. The correction procedure does effectively remove the unwanted influence of in-plane field components with no more than 5 iteration steps required. The optimal images are obtained with only 3–5 iterations. Due to the internal LabVIEW FFT routine, the artefacts in the form of fringes appear in the field and current mapping after >5 iteration steps, which, if done with a FFT routine having a smaller binning size, should not occur up to a large number of iteration steps.

A possible improvement to the presented in-plane correction technique might be to consider the magnetic anisotropy field as non-constant across the sample due to its micro-geometry or to account for these in-plane fields in a different way such as by deriving a modified Biot-Savart inversion procedure that avoids iterative calculations.[34]

## ACKNOWLEDGMENTS

The authors would like to thank F. Laviano for reading the manuscript and the fruitful, critical remarks. The authors have also benefited from discussions with I. Rudnev and M. Petasecca.

This work was supported by the Australian Research Council via A. V. Pan's Discovery Project No. DP110100398, by the Faculty of Engineering and Information Sciences (University of Wollongong, Australia), and the Australian Government Research Training Program (RTP) Scholarship.

## APPENDIX A: PROOF OF THE FOURIER TRANSFORM OF 1D INTEGRAL KERNEL

Proof of Eq. (10)
Consider

$$\mathcal{F}(e^{-h|x'|}) = \int_{-\infty}^{\infty} e^{-h|x'|} e^{-ik_x x'} dx'$$

$$= \int_{-\infty}^{0} e^{(h-ik_x)x'} dx' + \int_{0}^{\infty} e^{-(h+ik_x)x'} dx'$$

$$= \frac{2h}{h^2 + k_x^2}$$

$$\Longleftrightarrow e^{-h|x'|} = \mathcal{F}^{-1}\left(\frac{2h}{h^2 + k_x^2}\right) \text{ By the inversion theorem}$$

$$= \frac{1}{2\pi} \int_{-\infty}^{\infty} \frac{2h}{h^2 + k_x^2} e^{ik_x x'} dk_x$$

$$\Longleftrightarrow \frac{\pi}{h} e^{-h|k_x|} = \int_{-\infty}^{\infty} \frac{e^{-ik_x x'}}{h^2 + x'^2} dx'. \quad \text{(A1)}$$

Now, swapping the roles of $x'$ and $k_x$ and letting $k_x \to -k_x$ gives

$$\frac{\partial}{\partial k_x}\left(\frac{\pi}{h} e^{-h|k_x|}\right) = \frac{\partial}{\partial k_x}\left(\int_{-\infty}^{\infty} \frac{e^{-ik_x x'}}{h^2 + x'^2} dx'\right)$$

$$\Longleftrightarrow -i\pi \operatorname{sgn}(k_x) e^{-h|k_x|} = \int_{-\infty}^{\infty} \frac{x'}{h^2 + x'^2} e^{-ik_x x'} dx'.$$

Hence,

$$\mathcal{F}\left(\frac{x'}{x'^2 + h^2}\right) = \tilde{K}(k_x) = -i\pi \operatorname{sgn}(k_x) e^{-h|k_x|}. \quad \text{(A2)}$$

Q.E.D.

## APPENDIX B: PROOF OF THE FOURIER TRANSFORM OF 2D INTEGRAL KERNEL

Proof of the claim (25)
To prove the claim (25), first notice that the integral kernel

$$K(k_x, k_y) = k e^{-k(h-z')}, \quad \text{(B1)}$$

is radially symmetric. Hence, following the theorem of Hankel transforms, the co-ordinate system is changed using the following substitutions:

$$r = \sqrt{x^2 + y^2}, \quad \text{(B2)}$$

$$k_x = k \cos\theta, \, k_y = k \sin\theta \Rightarrow k = \sqrt{k_x^2 + k_y^2}, \quad \text{(B3)}$$

where $k, r \in [0, \infty)$, and $\theta \in [0, 2\pi]$. The standard change of double integrals to polar coordinates gives $dk_x dk_y = k\,dk\,d\theta$ and by writing $z = h - z'$ for simplicity, the right-hand side of Eq. (25) becomes

$$\frac{1}{2\pi}\int_0^{\infty}\int_0^{2\pi} k e^{-kz} e^{ikr\cos\theta} k\,dk\,d\theta = \frac{1}{2\pi}\int_0^{\infty} k^2 e^{-kz}\int_0^{2\pi} e^{ikr\cos\theta} dk\,d\theta. \quad \text{(B4)}$$

The Bessel function integral relation is then applied to the inner integral

$$\int_0^{2\pi} e^{ia\cos\theta} d\theta = 2\pi J_0(a), \quad \text{(B5)}$$

where $J_0(a)$ is the Bessel function of the first kind of order 0. Therefore

$$\frac{1}{2\pi}\int_0^{\infty}\int_0^{2\pi} k e^{-kz} e^{ikr\cos\theta} k\,dk\,d\theta = \int_0^{\infty} k^2 e^{-kz} J_0(kr) dk. \quad \text{(B6)}$$

Note, this integral represents the Laplace transform of the Bessel function multiplied by $k^2$ and multiplied in the argument by $k$. The integral can therefore be computed by finding the Laplace transform of the differential equation that defines this Bessel function, then using a frequency shift and a domain shift to account for the argument of $kr$ and the multiplication by $k^2$, respectively,

$$\int_0^{\infty} e^{-kz} J_0(k) dk = \mathcal{L}[J_0(k)]. \quad \text{(B7)}$$

As stated above, the right-hand side of Eq. (B7) is equivalent to computing the Laplace transform of the differential equation whose solution is $J_0(k)$

$$\mathcal{L}[x(y'' + y) + y'] = \mathcal{L}(0), \quad (B8)$$

since this is the equation that defines the Bessel function. The linearity and differentiation properties of the Laplace transform and normalisation properties of the Bessel functions are used to obtain

$$\mathcal{L}(J_0)(z) = \frac{1}{\sqrt{1+z^2}}, \quad (B9)$$

which is the well-known result for the Laplace transform of the Bessel function of the first kind of order 0.

Since the integrand involved the Laplace transform of the function $k^2 J_0(rk)$, time scaling, and general frequency domain differentiation properties of the Laplace transform can be used to give

$$\begin{aligned}\mathcal{L}[k^2 J_0(kr)] &= \frac{1}{r}\frac{d^2}{dz^2}\left[\mathcal{L}(J_0)\left(\frac{z}{r}\right)\right] \\ &= \frac{1}{r}\frac{d^2}{dz^2}\left(\frac{r}{\sqrt{r^2+z^2}}\right) \\ &= \frac{2z^2 - r^2}{(r^2+z^2)^{\frac{5}{2}}}.\end{aligned} \quad (B10)$$

Hence

$$\therefore \int_0^\infty e^{-k(h-z')} k^2 J_0(kr) dk = \frac{2(h-z')^2 - x^2 - y^2}{[x^2 + y^2 + (h-z')^2]^{\frac{5}{2}}}, \quad (B11)$$

verifying Eq. (25) under the prescribed change of co-ordinates, which corresponds to the original claim.

## APPENDIX C: PROOF OF THE INTEGRAL REPRESENTATION OF THE BESSEL FUNCTION

Proof of the Integral Relation (B5).
First, let us consider the integral expression

$$\frac{1}{\pi}\int_{-1}^{1} e^{izt}(1-t^2)^{-\frac{1}{2}} dt. \quad (C1)$$

Replacing the exponential in the integrand by its Taylor series expression

$$\frac{1}{\pi}\sum_{n=0}^{\infty}\frac{(iz)^n}{n!}\int_{-1}^{1}(1-t^2)^{-\frac{1}{2}}t^n dt. \quad (C2)$$

If $n$ is an odd integer, the function in the integrand is odd, hence integrating it over symmetric limits results in 0. So without loss of generality, it is possible to define $n=2k$ for $k \in \mathbb{Z}$, then since the integrand is now an even function

$$\frac{2}{\pi}\sum_{k=0}^{\infty}\frac{(iz)^{2k}}{(2k)!}\int_{0}^{1}(1-t^2)^{-\frac{1}{2}}t^{2k} dt. \quad (C3)$$

Under the change of variable $u = t^2$

$$\frac{1}{\pi}\sum_{k=0}^{\infty}\frac{(iz)^{2k}}{(2k)!}\int_{0}^{1}(1-u)^{-\frac{1}{2}}u^{k-\frac{1}{2}} du. \quad (C4)$$

Recalling the definition of the Beta function

$$B(x,y) = \int_0^1 t^{x-1}(1-t)^{y-1} dt = \frac{\Gamma(x)\Gamma(y)}{\Gamma(x+y)}, \quad (C5)$$

Eq. (C4) can be re-written as

$$\frac{1}{\sqrt{\pi}}\sum_{k=0}^{\infty}\frac{(iz)^{2k}}{(2k)!}\frac{\Gamma\left(k+\frac{1}{2}\right)}{k!}. \quad (C6)$$

Legendre's duplication formulae can be used to obtain an expression for $\Gamma\left(k+\frac{1}{2}\right)$

$$\Gamma\left(k+\frac{1}{2}\right) = \frac{\Gamma(2k)\sqrt{\pi}}{\Gamma(k)2^{2k-1}}. \quad (C7)$$

Substitution of the above relation into Eq. (C6) gives

$$\frac{1}{\pi}\int_{-1}^{1} e^{izt}(1-t^2)^{-\frac{1}{2}} dt = \frac{1}{\pi}\sum_{k=0}^{\infty}\frac{(-1)^k}{(k!)^2}\left(\frac{z}{2}\right)^{2k} = J_0(z) \quad (C8)$$

since the series expression derived is identical to the series expression for the Bessel function ($J_0(z)$) of the first kind of order 0.

Letting $t = \cos(\theta)$ in the equation above

$$\frac{1}{\pi}\int_0^\pi e^{iz\cos(\theta)} d\theta = J_0(z). \quad (C9)$$

Then, by the $\pi$-periodicity of the integrand, an extension of the limits of integration to $\theta \in [0, 2\pi]$ gives double the result in Eq. (C9)

$$\int_0^{2\pi} e^{iz\cos(\theta)} d\theta = 2\int_0^\pi e^{iz\cos(\theta)} d\theta. \quad (C10)$$

Finally, using Eq. (C10) to write Eq. (C9) in terms of the limits $\theta \in [0, 2\pi]$, the result is found

$$\frac{1}{2\pi}\int_0^{2\pi} e^{iz\cos\theta} d\theta = J_0(z). \quad (C11)$$